\documentclass[prl,showpacs,superscriptaddress,merge,preprint]{revtex4-1}

\usepackage[T1]{fontenc}
\usepackage[latin9]{inputenc}
\usepackage{float}
\usepackage{textcomp}
\usepackage{amsthm}
\usepackage{amsmath}
\usepackage{graphicx}
\usepackage{amssymb}
\usepackage{esint}

\newcommand{\ket}[1]{\left|#1\right\rangle}

\newcommand{\mean}[1]{\left\langle #1 \right\rangle}

\newcommand{\rp}{\right)}
\newcommand{\lp}{\left(}
\newcommand{\lcb}{\left\{}
\newcommand{\rcb}{\right\}}
\newcommand{\rsb}{\right]}
\newcommand{\lsb}{\left[}
\newcommand{\lbv}{\left|}
\newcommand{\rbv}{\right|}
\newcommand{\lvb}{\lbv}
\newcommand{\rvb}{\rbv}

\newcommand{\pref}[1]{(\ref{#1})}
\renewcommand{\eqref}[1]{Eq.~\pref{#1}}

\newcommand{\projector}{\Pi}
\newcommand{\proj}[1]{\projector_{#1}}

\begin{document}

\title{Quantum heating of a nonlinear resonator probed by a superconducting qubit}

\author{F.R. Ong}
 \affiliation{Quantronics group, Service de Physique de l'État Condens\'e
(CNRS URA 2464), IRAMIS, DSM, CEA-Saclay, 91191 Gif-sur-Yvette, France}
 \affiliation{Institute for Quantum Computing, Waterloo Institute for Nanotechnology, University of Waterloo, N2L 3G1,
Waterloo, Canada}
\author{M. Boissonneault}
 \affiliation{D\'epartement de Physique, Universit\'e de Sherbrooke, Sherbrooke, Qu\'ebec, Canada, J1K 2R1}
 \affiliation{Calcul Qu\'ebec, Universit\'e Laval, Qu\'ebec, Qu\'ebec G1V 0A6, Canada}
\author{F. Mallet}
  \affiliation{Laboratoire Pierre Aigrain, Ecole Normale Sup\'erieure, CNRS (UMR 8551), Universit\'e P. et M. Curie, 24, rue Lhomond, 75005 Paris, France}
\author{A.C. Doherty}
 \affiliation{Centre for Engineered Quantum Systems, School of Physics, The University of Sydney, Sydney, NSW 2006, Australia}
 \author{A. Blais}
 \affiliation{D\'epartement de Physique, Universit\'e de Sherbrooke, Sherbrooke, Qu\'ebec, Canada, J1K 2R1}
\author{D. Vion}
\author{D. Esteve}
\author{P. Bertet}
 \affiliation{Quantronics group, Service de Physique de l'État Condens\'e
(CNRS URA 2464), IRAMIS, DSM, CEA-Saclay, 91191 Gif-sur-Yvette, France}
 
 \email{patrice.bertet@cea.fr}

\date{\today}

\begin{abstract}
We measure the quantum fluctuations of a pumped nonlinear resonator, using a superconducting artificial atom as an in-situ probe. The qubit excitation spectrum gives access to the frequency and temperature of the intracavity field fluctuations. These are found to be in agreement with theoretical predictions; in particular we experimentally observe the phenomenon of quantum heating.
\end{abstract}

\pacs{Valid PACS appear here}

\maketitle


A resonator in which a nonlinear medium is inserted has rich dynamics when it is driven by an external pump field \cite{DykmanOUP}. In the case of a Kerr medium, the field amplitude inside such a Kerr nonlinear resonator (KNR) switches from a low to a high value when the pump reaches a certain threshold, a phenomenon known as bistability in optics \cite{LugiatoBistability1984} and bifurcation in the microwave domain \cite{Siddiqi2005}. This hysteretic transition between two dynamical states is a stochastic process triggered by fluctuations around the steady-state pumped oscillations. Given the potential applications for low-power all-optical logical elements such as switches and transistors \cite{nozaki_sub-femtojoule_2010}, a quantitative understanding of these fluctuations governing the sharpness of the transition, and therefore the device performance, is highly desirable \cite{dykman1988,vijay_invited_2009}. In a KNR operated in the quantum regime \cite{drummond_quantum_1980}, field fluctuations are mainly due to spontaneous parametric down-conversion (SPDC) \cite{harris_observation_1967,burnham_observation_1970,bouwmeester_experimental_1997} of pairs of photons at the pump frequency $\omega_{\rm p}$ into pairs of photons, one at the characteristic pumped-KNR frequency $\tilde{\omega}_{\rm c}$ and the other at the complementary idler frequency  $\omega_{\rm i}=2\omega_{\rm p}- \tilde{\omega}_{\rm c}$. The mode at $\tilde{\omega}_{\rm c}$ therefore acquires a thermal population with an effective temperature that depends on the pump frequency and power even when the electromagnetic bath to which it is coupled is at zero temperature, a central theoretical prediction known as quantum heating that remains to be tested~\cite{drummond_quantum_1980,marthaler_switching_2006,dykman_quantum_2011,andre_emission_2012}.

\begin{figure}[b!]
\begin{center}
\hspace{0mm}
\includegraphics[width=8.5cm,angle=0]{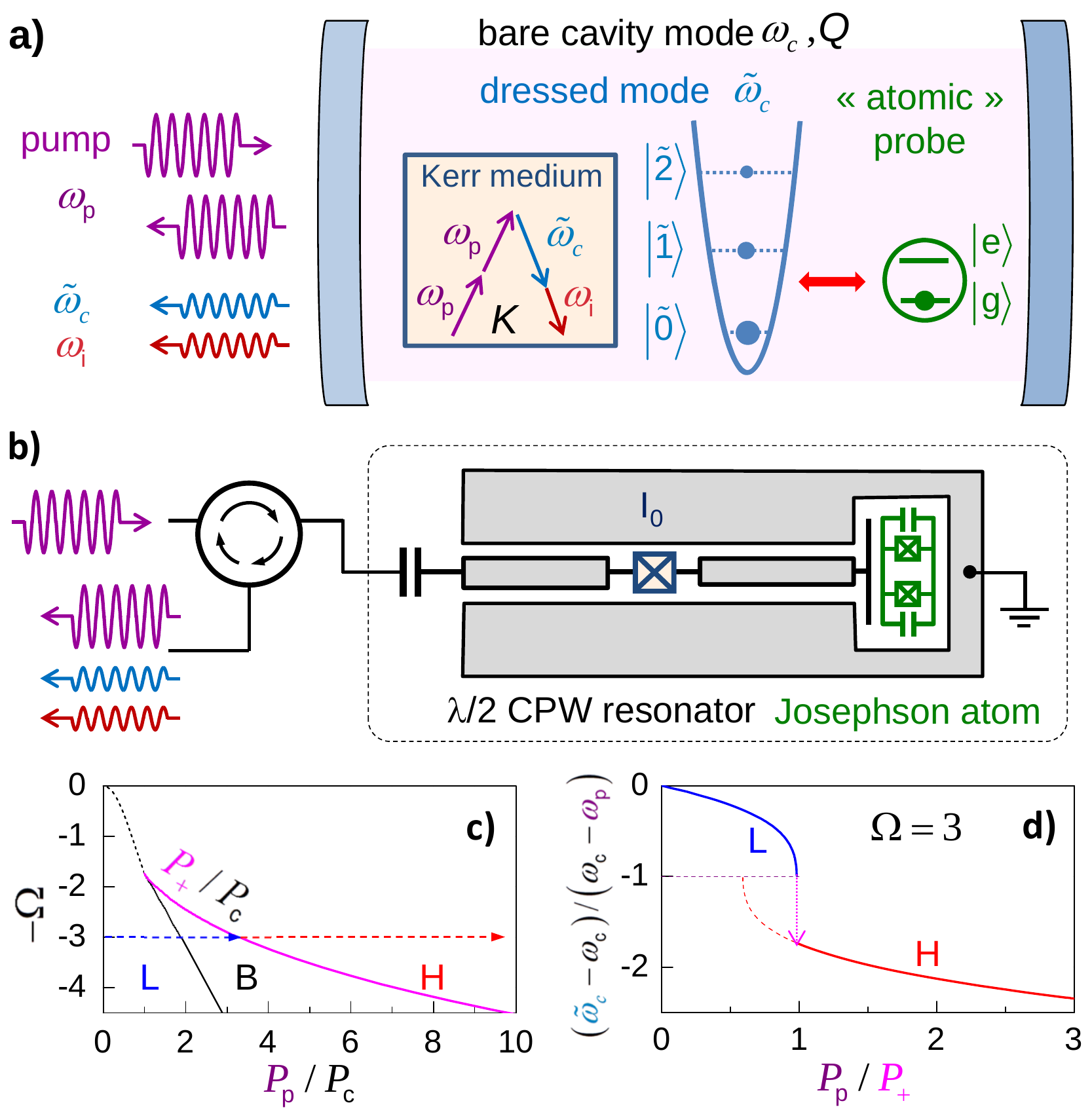}
\caption{(a) Optical analogue of the experiment: a Fabry-Pérot cavity with resonance frequency $\omega_{\rm c}$, quality factor $Q$, and Kerr non-linearity $K$, is pumped at frequency $\omega_{\rm p}$. Photons are produced at the cavity dressed mode frequency $\tilde{\omega}_{\rm c}$ and at the idler frequency $\omega_{\rm i}$ by SPDC of pump photon pairs. An in-situ atomic probe is used as an absolute spectrometer and thermometer to characterize this quantum heating of the dressed mode. (b) Actual experiment: the cavity is a microwave superconducting $\lambda/2$ coplanar waveguide resonator, the Kerr medium a Josephson junction, and the probe a transmon artificial atom. (c) Stability diagram of the resonator in the pump power $P_{\rm p}$ and frequency $\omega_{\rm p}$ plane (in reduced units, see text) : $L$, $H$, and $B$ denotes regions where the amplitude of the driven oscillations is low, high, and bistable, respectively. $P_+$ is the pump power at which the field switches from $L$ to $H$ when $P_{\rm p}$ is increased. (d) Dressed mode frequency $\tilde{\omega}_{\rm c}$ when increasing $P_{\rm p}$ at $\Omega=-3$ (along the arrows of panel c).
}
\label{fig1}
\end{center}
\end{figure}

Up to now, experiments in the optical or microwave domain have only measured the spectrum of the field radiated by a pumped KNR \cite{lahteenmaki_dynamical_2011}. In this work, we access the intra-resonator field fluctuations by inserting a two-level system (TLS) inside the KNR and using it as an absolute spectrometer and thermometer, as illustrated in Fig.~\ref{fig1}a. The experiment is performed in the microwave domain at millikelvin temperatures using superconducting Josephson circuits. The KNR is a coplanar waveguide resonator with an embedded Josephson junction~\cite{boaknin_dispersive_2007,ong_circuit_2011}, and the TLS is a transmon qubit~\cite{PhysRevA.76.042319} (see Fig.~\ref{fig1}b). By measuring the qubit absorption spectrum while pumping the KNR, we obtain a quantitative agreement with theoretical predictions on quantum heating.

We start with a rapid summary of the relevant theoretical results regarding the quantum fluctuations of a KNR - more details can be found in \cite{drummond_quantum_1980,dykman1988,marthaler_switching_2006,yurke_performance_2006,laflamme_quantum-limited_2011,andre_emission_2012,noteSM}. The KNR, driven by a field of power $P_{\rm p}$ and frequency $\omega_{\rm p}$, is modelled in the frame rotating at $\omega_{\rm p}$ by the Hamiltonian \linebreak $\bar{H}_{\mathrm c}/\hbar=\Delta_{\rm p} \bar{a}^\dagger \bar{a} + \frac{K}{2} \bar{a}^{\dagger 2} \bar{a}^2 + \frac{K'}{3} \bar{a}^{\dagger 3} \bar{a}^3 + (i \epsilon_{\rm p} \bar{a}^\dagger + \mathrm{H.c}).$ with $\Delta_{\rm p}=\omega_{\rm c} - \omega_{\rm p}$, $\omega_{\rm c}$ the KNR resonance frequency in its linear regime, $\bar{a}$ and $\bar{a}^\dagger$ the KNR field annihilation and creation operators, $\epsilon_{\rm p}=\sqrt{\kappa P_{\rm p} / \hbar \omega_{\rm p}}$ the driving amplitude, $\kappa$ the resonator energy damping rate, and $K$ and $K'$ the Kerr nonlinear constants derived from circuit parameters \cite{ong_circuit_2011,note_Kprime}. The steady-state solution for the dimensionless cavity field amplitude $\alpha$ is obtained from the corresponding master \linebreak equation yielding $ \left(\Omega \frac{\kappa}{2}+K|\alpha |^2+K' |\alpha|^4 -i\frac{\kappa}{2} \right)  \alpha=- i\epsilon_{\rm p}$ with $\Omega=2Q\Delta_{\rm p} / \omega_{\rm c}$ the reduced pump frequency \cite{noteSM,yurke_performance_2006}, and $Q=\omega_{\rm c} / \kappa $ the resonator quality factor. For $\Omega > \sqrt{3}$ and $P_{\rm p}$ larger than the critical power $P_{\rm c}=\frac{\kappa^2}{3\sqrt{3}|K|} \hbar \omega_{\rm p}$ this equation admits two stable solutions $\alpha_{\rm L,H}$ corresponding to metastable dynamical states $L$ and $H$ of respectively low- and high- amplitude \cite{yurke_performance_2006}. In this bistable and hysteretic regime, the transition from $L$ to $H$ occurs abruptly when ramping up the pump power at the bifurcation threshold $P_+(\Omega)$. The corresponding stability diagram is shown in Fig.~\ref{fig1}c.

We consider here the quantum fluctuations of the intra-resonator field around its steady-state value $\alpha$, in the regime where they are too weak to induce switching to the other dynamical state within the experiment duration. This justifies linearizing $\bar{H}_{\mathrm c}$ around $\alpha$, by writing $\bar{a}=\alpha + a$ and keeping only terms quadratic in $a$. Following~\cite{noteSM}, this linearized Hamiltonian can be diagonalized by introducing a new operator $\tilde{a}=\mu a + \nu a ^ \dagger$, and rewritten as $\tilde{H}_{\mathrm l}=\tilde{\Delta}_{\rm p} \tilde{a}^\dagger \tilde{a}$ with $\tilde{\Delta}_{\rm p} = \mathrm{sign} (A) \sqrt{B}$, $A=\Delta_{\rm p} + 2 K |\alpha|^2 + 3 K' |\alpha|^4$, and $B=A^2 - (K + 2 K' |\alpha|^2)^2 |\alpha|^4$. In the laboratory frame, intra-cavity field fluctuations are thus described as excitations of a harmonic dressed mode of resonance frequency $\tilde{\omega}_{\rm c} = \omega_{\rm p} + \tilde{\Delta}_{\rm p}$ that depends on the pump amplitude and frequency \cite{noteQuasiEnergy}. We note the eigenstates of this effective oscillator $\ket{\tilde{n}} = (\tilde{a}^\dagger)^n \ket{\tilde{0}}/\sqrt{n!}$\,. As shown in Fig.~\ref{fig1}d for the case where $K,K'<0$ and $\Delta_{\rm p} > 0$ as in our experiment, the dressed frequency $\tilde{\omega}_{\rm c}$ is equal to $\omega_{\rm c}$ at low pump power, then decreases when $P_{\rm p}$ is increased, reaching $\omega_{\rm p}$ when $P_{\rm p}=P_+$, which causes the field to jump to its high amplitude value and $\tilde{\omega}_{\rm c}$ correspondingly to jump well below $\omega_{\rm p}$. The dressed mode is damped at the same rate $\kappa$ as the KNR, but towards an equilibrium steady-state at a finite effective temperature $T_{\rm{eff}}$ corresponding to a mean number of excitations $\langle \tilde{n} \rangle$ equal to $|\nu|^2$, even if the bath physical temperature is zero~\cite{noteSM}. Physically this thermal population is caused by SPDC of pairs of pump photons at $\omega_{\rm p}$ into correlated photons at frequencies $\tilde{\omega}_{\rm c}$ and $\omega_{\rm i}=2\omega_{\rm p} - \tilde{\omega}_{\rm c}$, emitted in the dressed mode and in the measuring line respectively; the apparent thermal character of the intraresonator field is obtained when neglecting the correlations between the $\tilde{\omega}_{\rm c}$ and the $\omega_{\rm i}$ photons. Note that this analysis is only valid if $B > 0$, a condition verified sufficiently far from the bifurcation threshold, as is the case here.

\begin{figure}[t!]
\begin{center}
\hspace{0mm}
\includegraphics[width=8.5cm,angle=0]{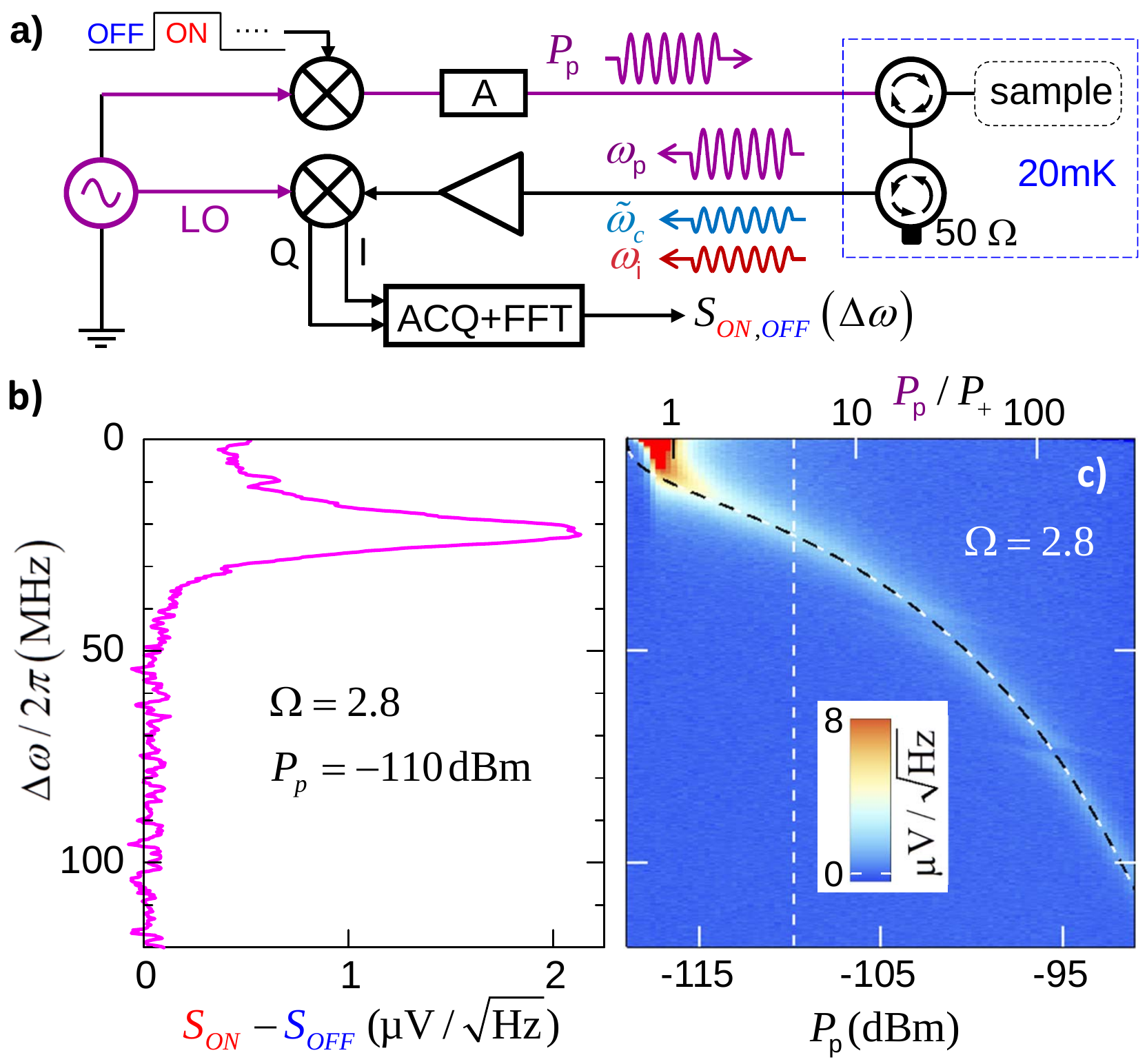}
\caption{Measurement of the field radiated by the KNR pumped at $\Omega=2.8$. (a) Experimental setup: the pump tone, sent through a line with attenuators ($A$), is switched on and off every 2.5\,ms; the reflected and radiated signals are amplified and mixed down at $\omega_{\rm p}$, yielding two quadratures $I(t)$ and $Q(t)$ that are digitized and Fourier transformed; the total voltage spectral densities $S_{\rm {ON,OFF}}(\Delta\omega=\left|\omega-\omega_p\right|)$ for the two pump states are then computed. (b) Radiated noise spectrum $S_{\rm {ON}}-S_{\rm {OFF}}$ for $P_{\rm p}=-110$\,dBm. (c) Radiated noise spectrum as a function of $P_{\rm p}$. The dashed curve is the calculated $\tilde{\Delta}_{\rm p}=\tilde{\omega}_{\rm c}-\omega_{\rm p}(P_{\rm p})$. The vertical dotted line indicates the cut shown on panel b.
}
\label{fig2}
\end{center}
\end{figure}

In our experimental test of these predictions, the KNR is a superconducting coplanar waveguide resonator including a Josephson junction with frequency $\omega_{\rm c}/2\pi=6.4535$\,GHz, Kerr constants $K/2\pi=-625$\,kHz and $K'/2\pi=-1.25$\,kHz, and damping rate $\kappa/2\pi=10$\,MHz (see Fig.~\ref{fig1}d)~\cite{ong_circuit_2011}. It is capacitively coupled with strength $g/2\pi = 44$\,MHz to a superconducting qubit of the transmon type with frequency $\omega_{\rm ge}/2\pi = 5.718$\,GHz. Because of the large qubit-resonator detuning, their interaction can be described in the so-called dispersive limit \cite{PhysRevA.69.062320}. Resonator and qubit can be driven by microwave pulses applied to the resonator input. The qubit can be readout in a single shot by driving the resonator close to its bifurcation threshold $P_+(\Omega)$ in order to map the qubit states $\left|g\right\rangle$ and $\left|e\right\rangle$ to the $L$ and $H$  states, respectively~\cite{mallet_single-shot_2009}.

A first method to investigate the field fluctuations of the KNR pumped at $\tilde{\omega}_{\rm c}$ is to measure the field radiated into the measurement line at frequencies $\tilde{\omega}_{\rm c}(P_{\rm p})$ and  $2\omega_{\rm p} - \tilde{\omega}_{\rm c}(P_{\rm p})$, as reported recently \cite{wilson_observation_2011,lahteenmaki_dynamical_2011}. Its noise spectrum, measured using the setup of Fig.~\ref{fig2}a~\cite{palacios-laloy_experimental_2010}, shows a peak at a $P_{\rm p}$-dependent frequency (see Figs.~\ref{fig2}b-c). All the sample parameters being known from earlier measurements~\cite{ong_circuit_2011}, the theoretical $\tilde{\omega}_{\rm c}(P_{\rm p})$ curve can be computed without adjustable parameters and is shown in Fig.~\ref{fig2}c. Except for avoided crossings at $\tilde{\Delta}_{\rm p} / 2\pi\approx40 $ and $70$\,MHz of unknown origin, the agreement is quantitative. This shows that the additional noise is indeed generated by the pumped KNR at $\tilde{\omega}_{\rm c}(P_{\rm p})$. However these measurements are unable to determine the effective temperature of the mode $\tilde{\omega}_{\rm c}$ from which the measured photons are leaking, which is the key quantity of the theory discussed above.

\begin{figure}[t!]
\begin{center}
\hspace{0mm}
\includegraphics[width=8.5cm,angle=0]{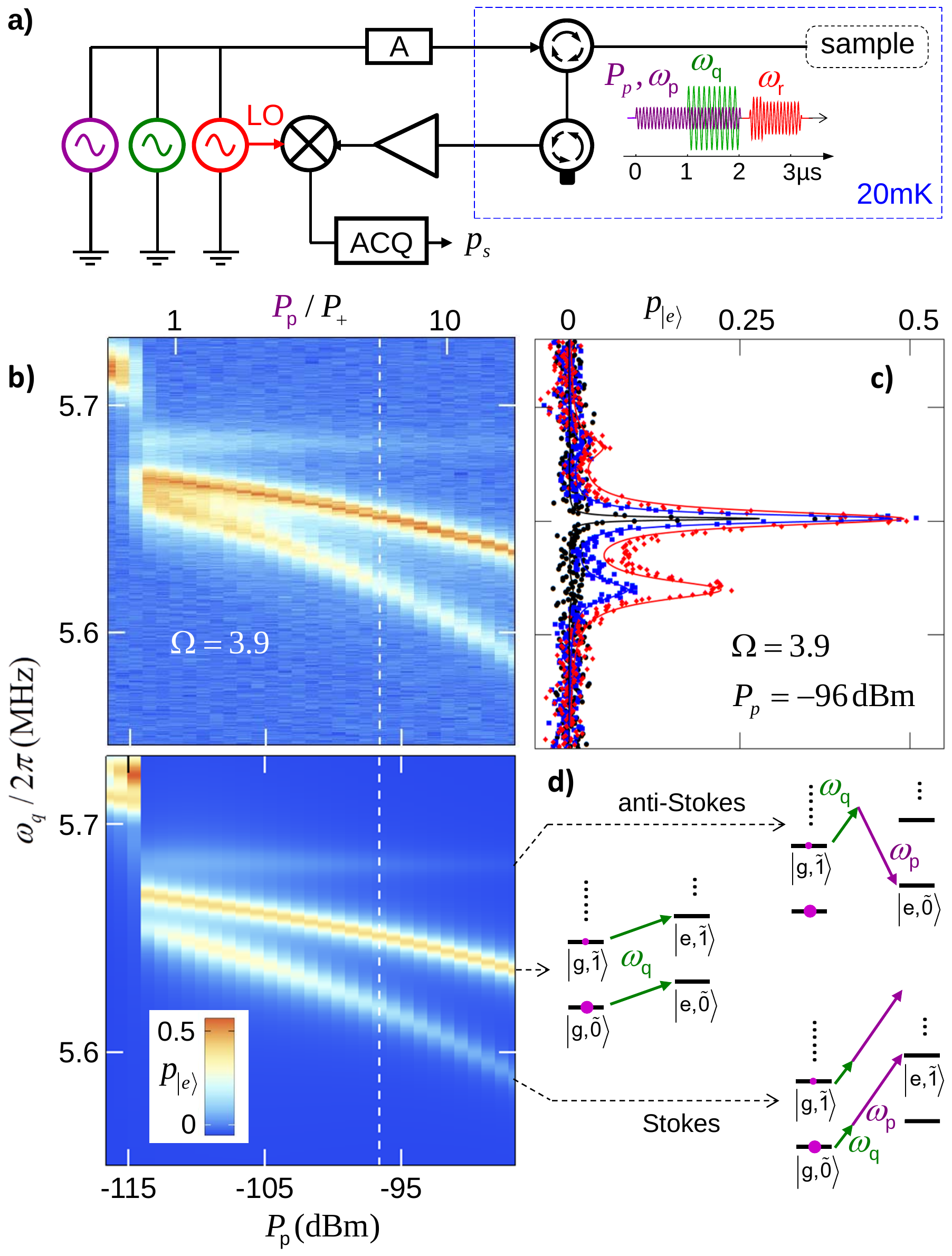}
\caption{Sideband spectroscopy of the qubit while pumping the resonator with varying power $P_{\rm p}$ at $\Omega=3.9$. (a) Experimental setup: a spectrocopic pulse is applied at $\omega_{\rm q}$ with power $P_{\rm q}$ during the pumping; the pump and spectrocopy tones are then switched off and the qubit is measured using a third pulse at $\omega_{\rm r}$ with a power close to the bifurcation threshold $P_+$. The measured phase of the reflected pulse yields the qubit state and repeating the sequence yields the qubit excited state probability $p_{\left|e\right\rangle}$.(b) Measured (top) and calculated (bottom) 2D plots showing the $p_{\left|e\right\rangle}(\omega_{\rm q})$ spectra as a function of $P_{\rm p}$ (bottom axis), also expressed in units of the bifurcation threshold $P_+$ (top axis). (c) Experimental (dots) and analytical (lines) spectra at a power $P_{\rm p}=-96$\,dBm (dashed lines in b), for powers $P_{\rm q}=$\, -131, -120 and -115\,dBm (from left to right). (d) Energy diagrams showing the transitions involved in the red sideband peak (Stokes), central peak, and blue sideband (anti-Stokes).
}
\label{fig3}
\end{center}
\end{figure}

A two-level system such as the transmon qubit is an ideal in-situ probe of quantum heating, because it acts as an absolute spectrometer and thermometer for a given quantum noise source~\cite{schoelkopf_qubits_2002}. We measure the temperature of the $\tilde{\omega}_{\rm c}$ mode while keeping the qubit and KNR far detuned by using a method called sideband spectroscopy, which is routinely used in ion-trapping experiments \cite{diedrich_laser_1989,Turchette2000} and has also been applied recently to mechanical oscillators \cite{safavi-naeini_observation_2012}. Indeed, starting from the system in state $\ket{g,\tilde{n}}$ it is possible to drive a transition to $\ket{e,\widetilde{n+1}}$ by irradiating the qubit at frequency $\omega_{\rm ge} + \tilde{\omega}_{\rm c}$ (so-called Stokes sideband), or to $\ket{e,\widetilde{n-1}}$ provided $n>0$ by irradiating the qubit at $|\omega_{\rm ge} - \tilde{\omega}_{\rm c}|$ (so-called anti-Stokes sideband)~\cite{chiorescu_coherent_2004,wallraff_sideband_2007,leek_using_2009}. In the case of a transmon qubit, these transitions need to be driven with two photons of arbitrary frequency provided their sum (resp. difference) satisfies the Stokes (resp. anti-Stokes) sideband resonance condition~\cite{Blais2007}. Given the sideband transition matrix element dependence on $n$, one can show that the average photon number $\langle \tilde{n} \rangle=1/[\exp(\hbar \tilde{\omega}_{\rm p}/k T_{\rm{eff}})-1]$ is equal to $r/(1-r)$ with $r\in\left[0,1\right]$ the anti-Stokes/Stokes sideband peak amplitude ratio \cite{Turchette2000}. Measurement of the qubit absorption spectrum, yielding $r$, therefore corresponds to a direct and absolute measurement of $T_{\rm{eff}}$. 

Sideband spectroscopy is performed using the setup shown in Fig.~\ref{fig3}a. Once a steady-state pump field is established in the resonator, a spectroscopy pulse is applied to the qubit at fixed power $P_{\rm q}$ and varying frequency $\omega_{\rm q}$. The pump tone provides one of the two photons needed to drive the sideband transitions, and the spectroscopy tone provides the second photon whenever $\omega_{\rm q}$ matches the Stokes (resp. anti-Stokes) sideband resonance condition $\omega_{\rm q} + \omega_{\rm p} = \omega_{\rm ge} + \tilde{\omega}_{\rm c}(P_{\rm p})$ (resp. $\omega_{\rm p} - \omega_{\rm q} = \tilde{\omega}_{\rm c}(P_{\rm p}) - \omega_{\rm ge}$). The experimental sequence ends by reading out the qubit state $200$\,ns after both pulses are switched off, long enough for the KNR field to decay but shorter than the qubit relaxation time $T_1 \approx 700$\,ns \cite{ong_circuit_2011}; repeating this sequence $\approx 10^4$ times yields the qubit excited state probability $p_{\left|e\right\rangle}$.

Typical data are shown in Fig.~\ref{fig3}c at $\Omega=3.9$ and $P_{\rm p}=-96$\,dBm such that the KNR is in the high oscillation amplitude state $H$. At low spectroscopy power $P_{\rm q}$, only one Lorentzian peak is visible, corresponding to the qubit frequency ac-Stark shifted by the steady-state intraresonator field with mean photon number $\langle n_{\rm H} \rangle = |\alpha_H|^2$ at $\omega_{\rm p}$ \cite{ong_circuit_2011,boissonneault_back-action_2012}. Increasing the spectroscopy power, we observe the appearance of two satellite peaks around $\omega_{\rm ge}$, with a separation of $31$\,MHz that closely matches the value of $\tilde{\Delta}_{\rm p}=\omega_{\rm p} - \tilde{\omega}_{\rm c}(P_{\rm p})$, already known without any adjustable parameters as explained above. When $P_{\rm p}$ is varied, this separation also quantitatively varies as expected from the $\tilde{\omega}_{\rm c}$ dependence on $P_{\rm p}$, as shown in Fig.~\ref{fig3}b. This establishes that the satellite peaks are indeed the sideband transitions. The anti-Stokes sideband being observable and of smaller amplitude than the Stokes sideband indicates that the temperature of the dressed mode is finite, as discussed in more details below.

To be more quantitative, we have performed a detailed theoretical analysis \cite{noteSM} of the coupled qubit-KNR system that will be presented elsewhere \cite{boissonneault_sidebands_2012}, and which yields analytical approximate expressions for the qubit sideband spectrum. The predictions, calculated with a global attenuation factor on the spectroscopy power $P_{\rm q}$ as the only adjustable parameter, are also shown in Fig.~\ref{fig3}b-c for different $P_{\rm q}$; they  agree quantitatively with the data. Since these calculations are done at zero bath temperature, this is a first clear indication that the population of the mode at $\tilde{\omega}_{\rm c}$ is only due to SPDC and is therefore of quantum origin.

\begin{figure}[t]
\begin{center}
\hspace{0mm}
\includegraphics[width=8.5cm,angle=0]{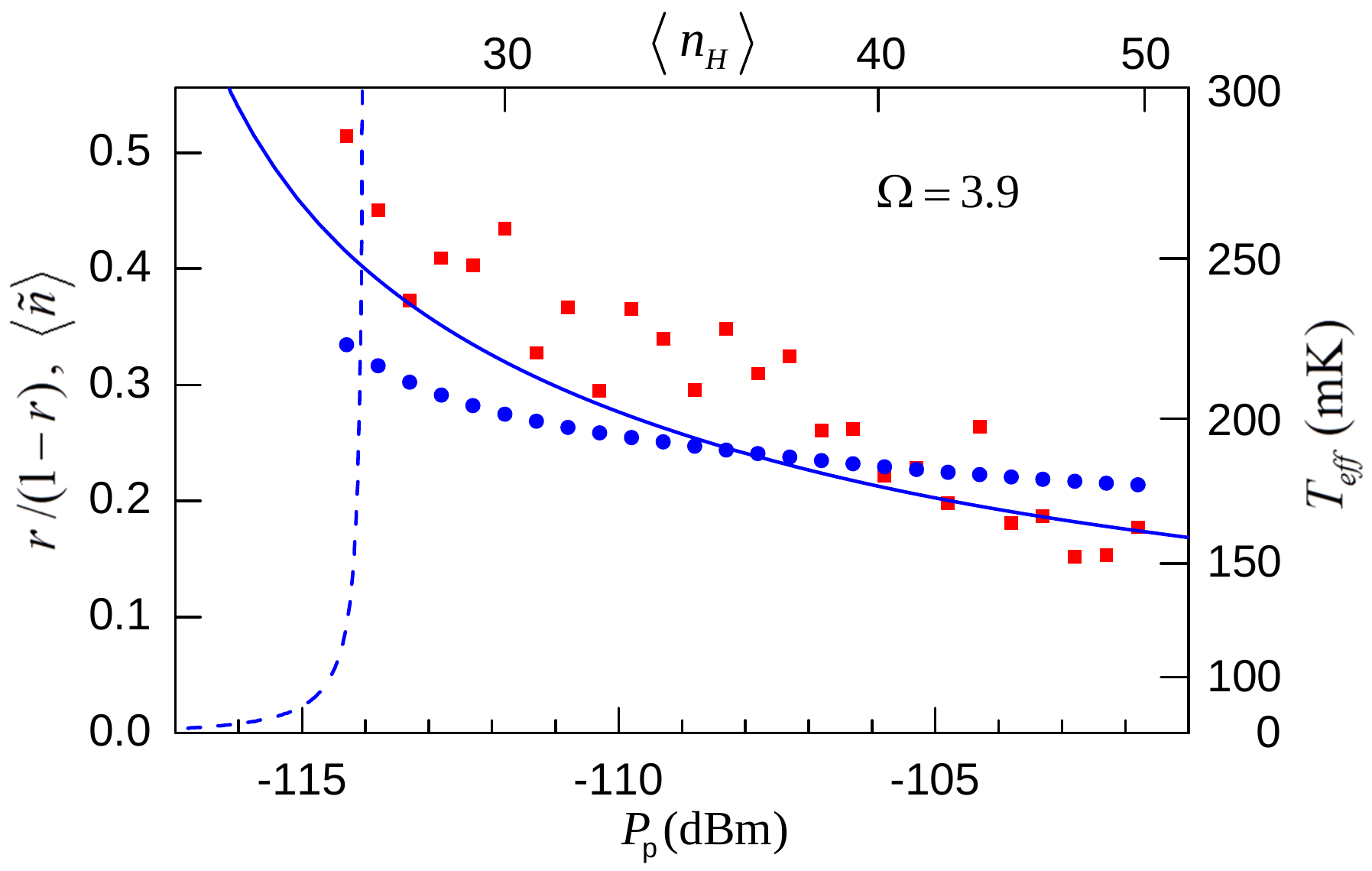}
\caption{Quantum heating of the pumped resonator mode $\tilde{\omega}_{\rm c}$ above bifurcation: Quantity $r/(1-r)$ with $r$ being the ratio of blue to red sideband peak heights deduced from the experimental (squares) and theoretical (dots) qubit spectra of Fig.~\ref{fig3}b, as a function of the pump power $P_{\rm p}$ (bottom axis). The calculated average number $\langle n_{\rm H} \rangle$ of photons in the pump mode is indicated on the top axis. In a simple theoretical picture $r/(1-r)$ is the average photon number $\langle \tilde{n} \rangle$ in the $\tilde{\omega}_{\rm c}$ mode, shown here as a dashed and a solid line  for the $L$ and $H$ resonator states. The corresponding effective temperature $T_{\rm{eff}}$ is indicated on the right axis.
}
\label{fig4}
\end{center}
\end{figure}

Assuming that the field in the dressed mode $\tilde{\omega}_{\rm c}$ has the statistics of a thermal field as predicted theoretically, we now extract an experimental occupation number of the dressed mode from the relative height of the two sideband peaks, and translate it into an effective temperature $T_{\rm eff}$. For this, each spectrum of Fig.~\ref{fig3}b is fitted to a sum of three Lorentzians of adjustable frequency, width and height, yielding the ratios $r$ of the anti-Stokes to Stokes sideband height. Figure \ref{fig4} shows the comparison between the experimental occupation number $r/(1-r)$ and $\langle \tilde{n} \rangle$ calculated without any adjustable parameter. The data agree fairly well with the prediction, demonstrating that the average photon number in the dressed mode is indeed $|\nu|^2$ as predicted by theory (see chapter $7$ in~\cite{DykmanOUP}). More precisely, we estimate that the residual thermal field at $\tilde{\omega}_{\rm c}$ cannot be responsible for more than $20\%$ of the observed signal. Our data show that the dressed mode occupation number and temperature are maximum near the bifurcation threshold $P_{\rm +}$ and decreases with increasing pumping power and occupation number $\langle n_{\rm H} \rangle$ at $\omega_{\rm p}$. This constitutes additional evidence that the measured thermal field is not due to a trivial heating caused by the microwave pulses. We also stress that these results do not rely on any calibration of the measurement lines or temperature. Finally, we also show in Fig.~\ref{fig4} the quantity $r/(r-1)$ derived from the analytical expression that yielded the theoretical spectra shown in Fig.~\ref{fig3}: it is in agreement with both the experimental data and the simple formula for the KNR effective temperature. Additional data can be found in the Supplementary Information \cite{noteSM} taken on another sample with different parameters, with which similar results have been obtained.

In conclusion, we have probed the quantum fluctuations of a pumped nonlinear resonator with an embedded superconducting qubit, bringing experimental evidence of quantum heating. Future directions include establishing the link between quantum heating and the switching rates at the bistability threshold~\cite{marthaler_switching_2006}, and testing the thermal character of the dressed resonator mode by performing its quantum state tomography with the qubit~\cite{hofheinz_synthesizing_2009}. In general our experiments demonstrate that detailed and quantitative tests of all theoretical predictions regarding nonlinear resonators in the quantum regime are enabled by the progress of superconducting circuits.

We acknowledge useful discussions with M.~Dykman and within the Quantronics group, as well as technical support from P.~S\'enat, P.-F.~Orfila, T.~David, and J.-C.~Tack. We acknowledge support from NSERC, the Alfred P. Sloan Foundation, CIFAR, the European project SOLID, the C'Nano project QUANTROCRYO, and the Australian Research Council via the Centre of Excellence in Engineered Quantum Systems (EQuS), project number CE110001013.

%

\begin{thebibliography}{38}
\bibitem [{not({\natexlab{b}})}]{note_Kprime}%
  \BibitemOpen
  \href@noop {} {} ({\natexlab{b}}),\ \bibinfo {note} {in order to obtain
  quantitative understanding of our experiments we need to include not only the
  standard Kerr nonlinearity term $\frac{K}{2} \bar{a}^{+2} \bar{a}^2$ but also
  the next order nonlinearity $\frac{K'}{3} \bar{a}^{+3} \bar{a}^3$. This term
  does not change the physics described in this article, hence we still speak
  of Kerr Nonlinear Resonator by convenience.}\BibitemShut {Stop}%
\bibitem [{not({\natexlab{c}})}]{noteQuasiEnergy}%
  \BibitemOpen
  \href@noop {} {} ({\natexlab{c}}),\ \bibinfo {note} {note that the separation
  between the energy levels of the effective resonator in the rotating frame
  $\hbar \tilde{\Delta}_{\rm p}$ is called quasi-energy by several
  authors.}\BibitemShut {Stop}%
\end{thebibliography}
%

\newpage

\section*{Supplementary Information for ``Quantum heating of a nonlinear resonator probed by a superconducting qubit''.}

\section{Theory of quantum fluctuations of a nonlinear resonator}

Our goal here is to give a self-contained account of the theory of quantum fluctuations of a nonlinear resonator. 

\subsection{Driven Kerr shifted cavity: linearized description}

Consider a Kerr nonlinear cavity with frequency $\omega_c$, Kerr coefficients $K$ and $K'$, and driven by
a field at frequency $\omega_p$ and amplitude $\epsilon_p$. The Hamiltonian describing the driven cavity
is
\begin{equation}
  \bar{H}_c=\hbar \omega_c \bar{a}^\dagger \bar{a} +\hbar \frac{K}{2} \bar{a}^{\dagger 2}  \bar{a}^2 +\hbar \frac{K'}{3} \bar{a}^{\dagger 3}  \bar{a}^3 +2 \hbar \epsilon_p \sin(\omega_p t) (\bar{a}+\bar{a}^\dagger)
\end{equation}

with $\bar{a}$ and $\bar{a}^\dagger$ the KNR field annihilation and creation operators.

In a frame rotating at $\omega_p$ and under the
rotating wave approximation, we find 
\begin{equation}
  H_c=\hbar \Delta_p \bar{a}^\dagger \bar{a} +\hbar \frac{K}{2} \bar{a}^{\dagger 2}  \bar{a}^2 +\hbar \frac{K'}{3} \bar{a}^{\dagger 3}  \bar{a}^3   +\hbar (-i\epsilon_p^*\bar{a}+i\epsilon_p \bar{a}^\dagger),
\end{equation}
where we have defined $\Delta_p=\omega_c-\omega_p$. Allowing the drive
amplitude $\epsilon_p$ to be complex in this equation allows for different
choices of drive phase. Including cavity damping $\kappa=\omega_c/Q$ ($Q$ being the cavity quality factor), the master equation in this frame is then
\begin{equation}
\label{eq:mastereq}
  \dot{\rho} = -i[H_c,\rho]/\hbar+ \kappa \mathcal{D}[\bar{a}]\rho,
\end{equation}
where
\begin{equation}
  \mathcal{D}[c]\rho=c\rho c^\dagger -c^\dagger c \rho/2-\rho c^\dagger c/2.
\end{equation}

In the limit in which the linearization is appropriate, the cavity is not too far from being in a coherent state. It is then useful to
define new operators $a,a^\dagger$ for the cavity such that
\begin{equation}
\label{eq:alpha}
  \bar{a}=\alpha +a,
\end{equation}
where $\alpha$ is a complex number which will be shown later to correspond to the coherent state
amplitude in the resonator. The operator $a$ describes fluctuations about
this coherent state, and the contribution to the photon number from
this state should be small compared to $|\alpha|^2$ for the
linearization to be a valid approximation. 

Making this replacement in $H_c$, we find the linearized Hamiltonian
\begin{equation}
  H_l=\hbar \left(\Delta + 2 K |\alpha|^2 + 3 K' |\alpha|^4 \right) a^\dagger a + \hbar \left( \frac{K}{2} \alpha^{*2} + K' |\alpha| ^2 \alpha^{*2} \right) a^2 + \hbar \left( \frac{K}{2} \alpha^{2} + K' |\alpha| ^2 \alpha^{2} \right) a^{\dagger 2},
\end{equation}
where the value of $\alpha$ is chosen such that all terms linear in
$a$, including those arising from the dissipation, cancel (see below). We have
also disregarded terms that involve higher powers of $a$ and
$a^\dagger$ on the grounds that the effects of these terms will be
small since they involve lower powers of $\alpha$ which is assumed to
be large compared to one.

\subsection{Bistable steady state amplitudes}

The value of $\alpha$ is found by requiring that, after substituting $\bar{a}=\alpha+a$ (Eq.~\ref{eq:alpha}) in the master equation Eq.\ref{eq:mastereq}, the terms proportional to $a$ have a zero coefficient. 
We first give the term linear in $a$ originating from the Hamiltonian part of the master equation:
\begin{equation}
 \hbar\left(-i\epsilon_p^*+\Delta \alpha^* + K |\alpha|^2 \alpha + K' |\alpha|^4 \alpha \right) .
\end{equation}
In the dissipative term of the master equation, the substitution Eq.~\ref{eq:alpha} leads to
\begin{equation}
  \mathcal{D}[\bar{a}]\rho=\mathcal{D}[a]\rho -i [i\hbar \frac{\kappa}{2}(\alpha^*a-\alpha a^\dagger),\rho]/\hbar,
\end{equation}
the last term of which contributes to the prefactor of $a$. Gathering all the terms linear in $a$ leads to $\alpha$ verifying
\begin{equation}
  i \epsilon_p^*=\Delta \alpha^* + K|\alpha|^2 \alpha^* + K'|\alpha|^4 \alpha^*  +i \frac{\kappa}{2} \alpha^*.
\end{equation}

Introducing the reduced variable $\Omega=2 Q \Delta_{\rm p} / \omega_{\rm c}$ this can be rewritten

\begin{equation}
  -i \epsilon_p= \left(\Omega \frac{\kappa}{2}  + K |\alpha |^2 + K' |\alpha |^4 - i \frac{\kappa}{2}  \right) \alpha,
\end{equation}

as given in the main text.

\subsection{Diagonalizing the linearized Hamiltonian}

In order to understand the linearized spectrum of the cavity we 
diagonalize the linearized Hamiltonian

\begin{equation}
   H_l=\hbar \left(\Delta + 2 K |\alpha|^2 + 3 K' |\alpha|^4 \right) a^\dagger a + \hbar \left( \frac{K}{2} \alpha^{*2} + K' |\alpha| ^2 \alpha^{*2} \right) a^2 + \hbar \left( \frac{K}{2} \alpha^{2} + K' |\alpha| ^2 \alpha^{2} \right) a^{+2}.
\end{equation}

We wish to diagonalize this by means of a Bogoliubov style squeezing
transformation by defining a new ``lowering'' operator
\begin{equation}
  \tilde{a}=\mu a + \nu a^\dagger
\end{equation}
and its corresponding conjugate
\begin{equation}
  \tilde{a}^\dagger=\mu^* a^\dagger + \nu^* a.
\end{equation}
Simple calculation shows that
\begin{equation}
  [\tilde{a},\tilde{a}^\dagger]=1
\end{equation}
if and only if the condition
\begin{equation}
  |\mu|^2-|\nu|^2=1
\end{equation}
holds. 

Consider now the Hamiltonian
\begin{equation}
  \tilde{H}_l = \hbar \tilde{\Delta}_{\rm p} \tilde{a}^\dagger\tilde{a}.
\end{equation}
We will show that for a suitable choice of $\mu$ and $\nu$ this
Hamiltonian corresponds to $H_l$ plus a constant.

Without loss of generality we will be able to choose $\mu$ to be real.
This constraint will hold identically if we define $\mu$ and
$\nu$ in terms of hyperbolic functions $\mu=\cosh r$,
$\nu=e^{2i\theta} \sinh r$. Again without loss of generality, we may also
require that $r\geq 0$ and hence $\sinh r\geq 0$.

The most straightforward way to proceed is to write the new
Hamiltonian in terms of the original operator and match coefficients :
\begin{eqnarray}
   \tilde{H}_l &=& \hbar \tilde{\Delta}_{\rm p} \left(\mu^* a^\dagger + \nu^* a\right)\left(\mu a + \nu a^\dagger \right) \\
&=& \hbar \tilde{\Delta}_{\rm p} \left( |\mu|^2 + |\nu|^2 \right) a^\dagger a + \hbar \tilde{\Delta}_{\rm p} \left( \mu^*\nu a^{\dagger 2} +\mu\nu^* a^2\right) +\hbar \tilde{\Delta}_{\rm p} |\nu|^2.
\end{eqnarray}
We can see that this is equal to $H_l$ up to a constant as long as we
can choose
\begin{eqnarray}
  \tilde{\Delta}_{\rm p} \cosh (2r) &=& \Delta + 2K|\alpha|^2 + 3K'|\alpha|^4, \\
  \frac{1}{2}\tilde{\Delta}_{\rm p} \sinh (2r) e^{-2i\theta} &=& \left( \frac{K}{2} + K'|\alpha|^2  \right) \alpha^{*2}.
\end{eqnarray}

Notice that the first equation determines the sign of $\tilde{\Delta}_{\rm p}$ since
$\cosh$ is positive: $\tilde{\Delta}_{\rm p}$ must have the same sign as $\Delta +
2K|\alpha|^2+ 3K'|\alpha|^4$.

Looking at the second equation, we will define
$\alpha=|\alpha|e^{i\theta_c}$ and then we can conclude that $\exp
(-2i\theta)=\pm \exp (-2i\theta_c)$. Recalling that we have chosen
$r\geq 0$ implying that $\sinh r$ is positive, the plus sign is obtained when
$\tilde{\Delta}_{\rm p}$ has the same sign as $K$ and the minus sign otherwise. In
the former case we have $\theta=\theta_c$ and in the latter
$\theta=\theta_c+\pi/2$. In either case the squeezing axis rotates with the
phase of the cavity amplitude.

By taking the ratio of the absolute values of these equations we find
an expression for $r$
\begin{equation}
  \tanh 2r = \frac{|K + 2K' |\alpha|^2 | |\alpha|^2}{|\Delta + 2K|\alpha|^2 + 3K'|\alpha|^4|},
\end{equation}
while the magnitude of $\tilde{\Delta}_{\rm p}$ can be inferred from squaring and
substracting the two equations:

\begin{equation} 
\label{eq:DeltaTilde}
  \tilde{\Delta}_{\rm p}^{2}= (\Delta + 2K|\alpha|^2 + 3K'|\alpha|^4)^2 - (K + 2K' |\alpha|^2)^2|\alpha|^4.
\end{equation}

The regime of interest for us is where $(\Delta + 2K|\alpha|^2 + 3K'|\alpha|^4)^2 - (K + 2K' |\alpha|^2)^2|\alpha|^4 > 0$
in which case $\tilde{\Delta}_{\rm p}$ is real and the dynamics of the KNR fluctuations can indeed be mapped onto those of a damped harmonic
oscillator of frequency $\tilde{\Delta}_{\rm p}$ in the rotating frame. This is the case in most of the KNR phase space, except very close to the bistability threshold, in which case the fluctuations are instead amplified and the system behaves as a parametric amplifier of large gain. In this work we only focus on the regime where $\tilde{\Delta}_{\rm p}$ is real.

In this regime we can thus interpret the system as a cavity of
frequency $\tilde{\omega}_c$ with $\tilde{\omega}_c = \tilde{\Delta}_{\rm p} + \omega_p$.

\subsection{Approximate master equation in the linearized description : quantum heating}

 In the regime where $\kappa \ll \tilde{\omega}_c$, we can move into a new interaction picture at
 frequency $\tilde{\omega}_c$. In this interaction picture, the Hamiltonian is
 zero and only the dissipative terms in the master equation
 contribute. We have so far written these terms only as a function of
 $a$ but, when we express them in terms of $\tilde{a}$ using the
 inverse transform mentioned above and move into the interaction
 picture, we find that some terms have a time dependence at frequency
 $2\tilde{\Delta}_{\rm p}$. In the limit $\tilde{\Delta}_{\rm p} \gg \kappa$ in which we are working (corresponding to the 
 resolved-sideband limit), one
 can make a rotating wave approximation and average these terms to
 zero. The resulting master equation is
 \begin{equation}
   \dot{\rho} = \kappa (|\nu|^2+1)\mathcal{D}[\tilde{a}]\rho
 +\kappa |\nu|^2\mathcal{D}[\tilde{a}^\dagger]\rho.
 \end{equation}

 This is the master equation for an oscillator coupled to a thermal
 bath with energy decay rate $\kappa$ and thermal expectation number
 $|\nu|^2$. The
 steady state of the cavity mode is thus a thermal state with $\langle
 \tilde{a}^\dagger\tilde{a}\rangle = |\nu|^2$. It may seem strange to
 have a thermal state as a steady state. Physically the quantum state
 in this interaction picture approaches a squeezed state with some
 periodic time dependence of the phase of the squeezing. This variation
 is so rapid that, as long as we respect the regime of validity of the
 rotating wave approximation, we can replace it with a time averaged
 state which is thermal. This phenomenon has been coined "`quantum heating"' 
 \cite{marthaler_switching_2006-1,dykman_quantum_2011-1,andre_emission_2012-1}.

\section{Analytical formula for the sideband spectrum} 

\newcommand{\inp}[1]{\lp #1 \rp}
\newcommand{\insb}[1]{\lsb #1 \rsb}
\newcommand{\incb}[1]{\lcb #1 \rcb}
\newcommand{\invb}[1]{\lvb #1 \rvb}

The analytical results presented in Fig.~3 of the paper were obtained using a more complete theory than that presented above. This theory is based on the one developped in Ref.~\cite{boissonneault_back-action_2012-1}, and will be developped further in Ref.~\cite{boissonneault_sidebands_2012-1}. It includes effects from the resonator on the qubit such as Purcell decay, measurement-induced dephasing~\cite{ong_circuit_2011-1} and dressed dephasing~\cite{PhysRevA.77.060305-1} and decay, as well as Lamb and ac-Stark shift of the qubit transition frequencies. It also includes effects from the qubit on the resonator, such as a qubit-state dependent pull of the resonator's frequency leading to different pointer states $\alpha_i$ of the resonator for each qubit state $|i\rangle$. The theory starts from a master equation description of a qubit coupled to a nonlinear resonator driven both with a pump $\epsilon_p$ and spectroscopy $\epsilon_s$ drives. Through successive transformations~\cite{boissonneault_sidebands_2012-1}, we transform the master equation into a form in which the resonator can be adiabatically eliminated, yielding a reduced qubit master equation given by
\begin{equation}
	\dot \rho_q = -i [H,\rho_q] + \tilde\gamma_\downarrow \mathcal{D}[\sigma_-]\rho_q + \tilde\gamma_\uparrow\mathcal{D}[\sigma_+]\rho_q + \frac{\tilde\gamma_\varphi}{2}\mathcal{D}[\sigma_z]\rho_q,
\end{equation}
where
\begin{equation}
	H = \frac{\delta}{2}\sigma_z + g_0(\alpha_s\sigma_+ + \alpha_s^*\sigma_-).
\end{equation}
Here, $\alpha_s$ is the intra-resonator field created by the spectroscopy drive of amplitude $\epsilon_s$ at $\omega_s = \tilde\omega_{1,0} - \delta$, close to the qubit transition frequency. The dephasing rate $\tilde\gamma_\varphi$ includes the intrinsic qubit dephasing as well as the measurement-induced dephasing~\cite{boissonneault_sidebands_2012-1}. The up and down rates are given by
\begin{subequations}
	\begin{align}
		\tilde\gamma_\downarrow &= \gamma_\downarrow''' + |g_0 \alpha_s f(\alpha_i,\nu,\mu)|^2 \left[ \left( L(-\delta)+L(\delta)\right) |\nu|^2 + L(-\delta) \right] \\
		\label{eqn:gamma_uparrow}
		\tilde\gamma_\uparrow &= \gamma_\uparrow''' + |g_0 \alpha_s f(\alpha_i,\nu,\mu)|^2 \left[ \left( L(-\delta)+L(\delta)\right) |\nu|^2 + L(\delta) \right],
	\end{align}
\end{subequations}
where $\gamma_{\downarrow,\uparrow}'''$ include Purcell relaxation and dressed dephasing and are obtained in Ref.~\cite{boissonneault_back-action_2012-1} and $L(\delta)$ is a Lorentzian of full-width at half-maximum $\kappa$ given by
\begin{equation}
	L(\delta) \approx \frac{\kappa/2}{\kappa^2/4+(\tilde\Delta_p + \delta)^2}.
\end{equation}
In this equantion $\tilde{\Delta}_p$ is given by Eq.~\ref{eq:DeltaTilde} and $f(\alpha_i,\nu,\mu)$ is a unit-less function of the squeezing coefficients and of the distinguishability of the pointer states $\alpha_{1}-\alpha_{0}$. 

The stationnary solution of the reduced master equation is analytical and yields
\begin{equation}
	\label{eqn:mean_proj1_qubit}
	P(\ket{1}) = \frac{\mean{\proj{1,1}}_\mathrm{eq}\inp{\tilde\gamma_2^2+\delta^2} + 2\tilde\gamma_2\invb{g_0\alpha_{s,0}}^2/(\tilde\gamma_\uparrow+\tilde\gamma_\downarrow)}{\insb{\inp{\tilde\gamma_2^2+4\tilde\gamma_2\invb{g_0\alpha_{s,0}}^2/(\tilde\gamma_\uparrow+\tilde\gamma_\downarrow) }+\delta^2}},
\end{equation}
where
\begin{equation}
	\mean{\proj{1,1}}_\mathrm{eq} = \frac{\tilde\gamma_\uparrow}{\tilde\gamma_\uparrow+\tilde\gamma_\downarrow},
\end{equation}
and $\tilde\gamma_2 = \tilde\gamma_\varphi + (\tilde\gamma_\uparrow+\tilde\gamma_\downarrow)/2$. Equation~(\ref{eqn:mean_proj1_qubit}) was used to plot the analytical data presented in Fig.~3 of the paper. 

In the low spectroscopy power regime, the effective rates $\tilde\gamma_{\uparrow,\downarrow,2}$ are independant of the spectroscopy frequency, and the spectrum yields a single line, centered at $\delta=0$, with a minimal width of $\tilde\gamma_2$. In the high spectroscopy power regime, $\alpha_s$ becomes significant and the central line is power-broadened. Moreover, the rates acquire a frequency-dependent structure. If we focus on the well-resolved sidebands limit, that is $\delta > \tilde\gamma_2$, the dominant term yields $P(\ket{1}) \approx \mean{\proj{1,1}}_\mathrm{eq}$. One can then easily compute the blue to red sidebands ratio of amplitude simply with
\begin{equation}
	\frac{\left.\mean{\proj{1,1}}_\mathrm{eq}\right|_{\delta=\tilde\Delta_p}}{\left.\mean{\proj{1,1}}_\mathrm{eq}\right|_{\delta=-\tilde\Delta_p}} = \frac{\left.\tilde\gamma_\uparrow\right|_{\delta=\tilde\Delta_p}}{\left.\tilde\gamma_\uparrow\right|_{\delta=-\tilde\Delta_p}} \approx \frac{\insb{L(-\tilde\Delta_p) + L(\tilde\Delta_p)} |\nu|^2 + L(\tilde\Delta_p)}{\insb{L(-\tilde\Delta_p) + L(\tilde\Delta_p)}|\nu|^2 + L(-\tilde\Delta_p)} \approx \frac{|\nu|^2}{|\nu|^2 + 1}.
\end{equation}
In the first approximation, we assumed that the up rate $\gamma_\uparrow'''$ is negligible, while in the second approximation, we assumed that $L(-\tilde\Delta_p) \gg L(\tilde\Delta_p)$, which should be the case in the well resolved sidebands limit. 
 
\section{Data from another sample}
 
To complement the data shown in Fig.~4 of the main text, we now present data showing sideband spectroscopy of a similar sample, although with different values of the parameters. For this sample B, $\omega_c/2\pi = 6.469$\,GHz, $Q=1040$, and $K/2\pi=1$\,MHz. $K$ is determined as explained in \cite{ong_circuit_2011-1} by fitting the frequency shift (AC Stark shift) of the qubit while the KNR is pumped. The pump frequency was chosen to be $\omega_p/2\pi = 6.427$\,GHz, corresponding to a significantly larger $\Omega = 13.5$ than in the main text. 

\begin{figure}[t]
\begin{center}
\hspace{0mm}
\includegraphics[width=8.5cm,angle=0]{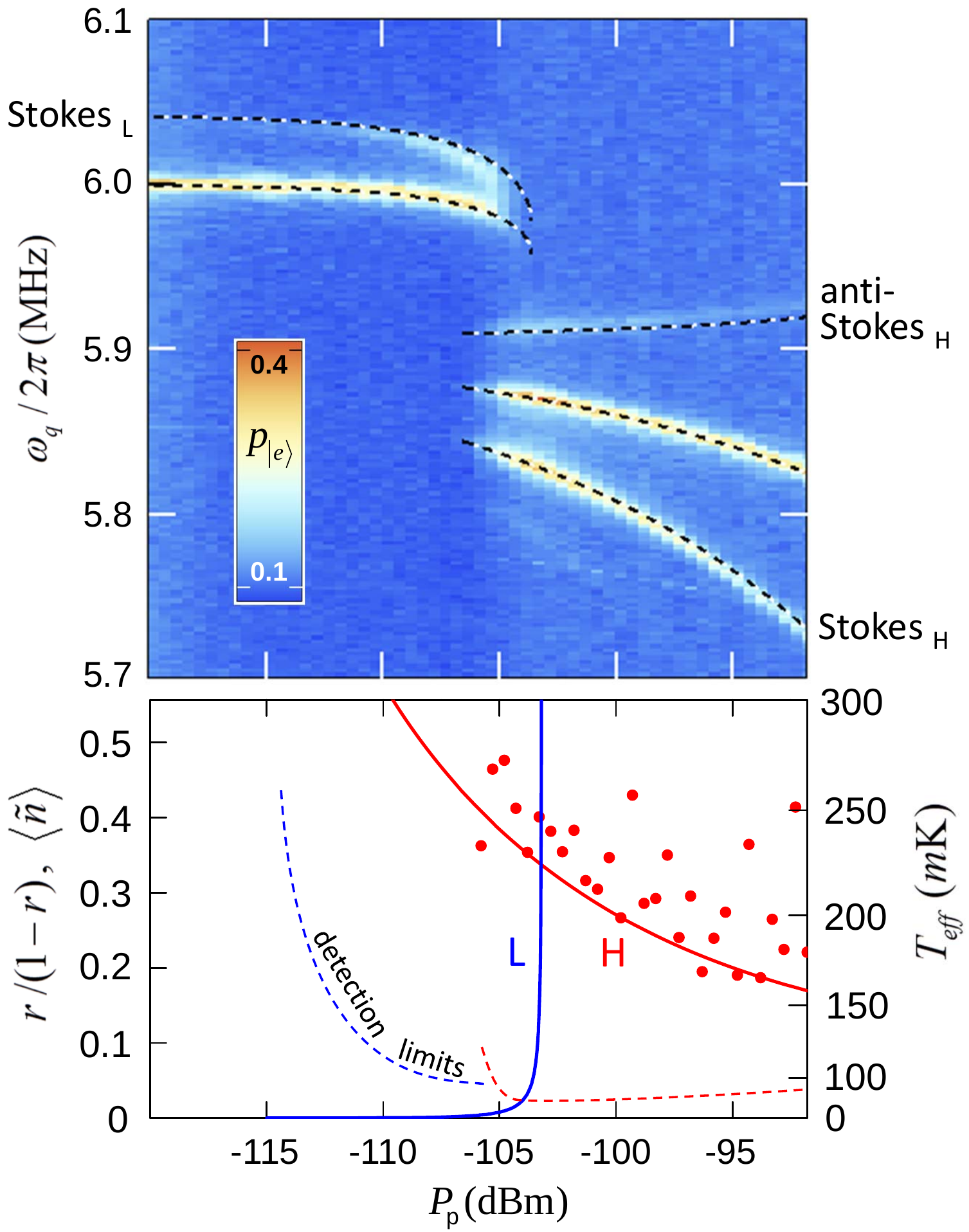}
\caption{(a) (color plot) Sideband spectroscopy of sample B on both sides of the bistability threshold, i.e. for the resonator in the $L$ state ($P_{\rm p}<-105$\,dBm) and in the $H$ state ($P_{\rm p}>-105$\,dBm). The expected frequency of the AC Stark-shifted main qubit peak as well as the Stokes and anti-Stokes sidebands are shown as  dashed white and black lines. (b) (red dots) Measured $r/(1-r)$, with $r$ being the ratio of the anti-Stokes to Stokes sideband peak height. Data points are shown only above the threshold, because no measurable anti-Stokes peak is found in our data below threshold. We show instead our detection limit, defined as $nf/(1-nf)$, with $nf$ the ratio of the standard deviation in $p_{| e \rangle}$ to the Stokes sideband peak height, above (resp. below) threshold, as a blue (resp. red) dashed line. The detection limit depends on the pump power because the Stokes sideband height does. The value of $\langle \tilde{n} \rangle$ calculated without adjustable parameter is shown as full line above (red) and below (blue) the threshold.
}
\label{figS1}
\end{center}
\end{figure}

Sideband spectroscopy was performed similarly to what is reported in the main text. Data are shown in Fig.\ref{figS1}, around the bistability threshold. Above the bistability threshold the data are similar to the data presented in Fig.3 of the main text. Note the additional presence of a faint fourth line at $\omega_{ge} + 2 (\tilde{\omega}_{\rm c} (P_{\rm p}) - \omega_{\rm p}) $, that corresponds to a transition from $\ket{g,\tilde{n}}$ to $\ket{e,\widetilde{n+2}}$, and of a narrow peak that corresponds to the two-photon excitation of the qubit from state $\ket{g}$ to its second excited state, AC-Stark shifted by the field inside the KNR. Below threshold (resonator in state $L$), the Stokes sideband is also well resolved and visible. This was not the case in the sample discussed in the main text because the pump frequency was much closer to the KNR frequency in these data, resulting in a smaller separation between the Stokes sideband below trheshold and the main qubit peak (since well below the threshold $\tilde{\Delta}_{\rm p} \approx \omega_c - \omega_p$), which made the two lines indistinguishable. Quite remarkably, even though the Stokes sideband is clearly visible below the threshold, the anti-Stokes sideband is invisible within our detection limit (see Fig.\ref{figS1}). Since the ratio between Stokes and anti-Stokes peaks directly yields the mean photon number $\langle \tilde{n} \rangle$, this sets an upper bound on the mean photon number $\langle \tilde{n} \rangle$ present in the mode at $\tilde{\omega}_{\rm c}$ which is shown in Fig.\ref{figS1} for various pump powers (the dependence on pump power is simply due to the fact that the height of the Stokes peak also diminishes at small pump power, whereas the noise in the data is constant). In particular, this allows us to establish that any residual photon numbers that might be present at $\tilde{\omega}_{\rm c}$ due to improper thermalization or filtering is below the lower point of the dashed blue line in Fig.\ref{figS1}, namely $<0.04$. This is again strong evidence that the much larger thermal photon numbers observed above threshold are genuinely of quantum origin.

Another noteworthy feature is that the Stokes and anti-Stokes sideband position with respect to the qubit frequency $\omega_q$ are inverted at the bistability threshold, as expected from the sign change of $\tilde{\Delta}_{\rm p}$ (see Fig.1 in the main text). Finally, the effective temperature extracted from Fig.\ref{figS1}a is shown in Fig.\ref{figS1}b and is again in fair agreement with the prediction $|\nu|^2$ calculated without adjustable parameter, both below and above the threshold.

\end{document}